\documentclass[12pt,preprint]{aastex}

\newcommand{\myemail}{palmeri@umh.ac.be}

\shorttitle{Radiative and Auger data for Ni K lines}
\shortauthors{Palmeri et al.}

\begin{document}


\title{Radiative and Auger decay data for modelling nickel K lines}


\author{P. Palmeri,}
\affil{Astrophysique et Spectroscopie, Universit\'e de Mons-Hainaut, B-7000 Mons, Belgium}
\email{\myemail}

\author{P. Quinet,}
\affil{Astrophysique et Spectroscopie, Universit\'e de Mons-Hainaut, B-7000 Mons, Belgium, and
\\ IPNAS, Sart Tilman B15, Universit\'e de Li\`ege, B-4000 Li\`ege, Belgium}
\email{quinet@umh.ac.be}

\author{C. Mendoza,}
\affil{Centro de F\'isica, IVIC, Caracas 1020A, Venezuela}
\email{claudio@ivic.ve}

\author{M.A. Bautista,}
\affil{Department of Physics, Virginia Polytechnic and State University, Blacksburg, VA 24061}
\email{bautista@vt.edu}

\author{J. Garc\'ia,}
\affil{IACS-Department of Physics, The Catholic University of America, Washington, DC 20064}
\email{javier@milkyway.gsfc.nasa.gov}

\author{M.C. Witthoeft and T.R. Kallman}
\affil{NASA Goddard Space Flight Center, Greenbelt, MD 20771}
\email{michael.c.witthoeft@nasa.gov; timothy.r.kallman@nasa.gov}


%\altaffiltext{1}{Visiting Astronomer, Cerro Tololo Inter-American Observatory.
%CTIO is operated by AURA, Inc.\ under contract to the National Science


\begin{abstract}
Radiative and Auger decay data have been calculated for modelling
the K lines in ions of the nickel isonuclear sequence, from Ni$^+$
up to Ni$^{27+}$. Level energies, transition wavelengths,
radiative transition probabilities, and radiative and Auger widths
have been determined using Cowan's Hartree--Fock with Relativistic
corrections (HFR) method. Auger widths for the third-row ions
(Ni$^+$--Ni$^{10+}$) have been computed using single-configuration
average (SCA) compact formulae. Results are compared with data
sets computed with the AUTOSTRUCTURE and MCDF atomic structure
codes and with available experimental and theoretical values,
mainly in highly ionized ions and in the solid state.
\end{abstract}





\clearpage

\section{Introduction}

The nickel K lines have arguable diagnostic potential in X-ray
astronomy. They reside in a relatively unconfused part of the
X-ray spectrum where they can be used to estimate quantities of
interest such as the redshift, temperature, abundance, and the
velocity of the emitting gas. Sensitive observations from the
currently active {\em Suzaku} satellite, and those to be expected
from future space missions, e.g. the {\em Constellation-X}
observatory, are bringing about further spectroscopic attention to
the Ni K features.

On the other hand, clear detections of nickel have been obtained
only in a few astrophysical X-ray spectra. This is undoubtedly due
to a combination of limited counting statistics at energies above
7 keV, where the nickel lines occur, and to poor spectral
resolution in this band which makes nickel $K\alpha$ and iron
$K\beta$ lines indistinguishable. The first detection of Ni
K$\alpha$ in an X-ray binary was reported by \cite{Sak99} in the
{\em ASCA} spectrum of Vela X-1. Fluorescence line energies and
intensities were measured mostly for neutral elements, including
nickel in eight ionization stages in spite of the uncertainties in
the line energy ($7.52^{+0.06}_{-0.05}$ keV). This detection was
later confirmed by \cite{Gol04} using three data sets obtained
with {\em Chandra/HETGS} during a binary orbit, representing the
orbital eclipse phases of $\phi=0$, $\phi=0.25$, and $\phi=0.5$.
They found relative strong Ni K$\alpha$ emission with an increase
in the line flux of a factor of 15 between the $\phi=0$ and
$\phi=0.5$ phases.

X-ray absorption features from Ni have been found in X-ray binary
systems. \cite{Boi04} presented evidence of K$\alpha$ absorption
in Ni~{\sc xxvii} at 7.82 keV in the {\em XMM-Newton} spectrum of
the low-mass X-ray binary (LMXB) XB 1916-053. In the case of the
LMXB GX 13+1, \cite{Sid02} came across two absorption features at
7.85 and 8.26 keV which they respectively identified as either
K$\beta$ lines from Fe~{\sc xxv} and Fe~{\sc xxvi} or K$\alpha$
from Ni~{\sc xxvii} and Ni~{\sc xxviii}.

Ni K$\alpha$ emission in an active galactic nucleus (AGN) was
first detected in the {\em XMM-Newton} X-ray spectra of the
Circinus galaxy \citep{Mol03}. Fe K$\alpha$, Fe K$\beta$, and Ni
K$\alpha$ were identified at 6.4, 7.058, and 7.472 keV,
respectively. From the Ni and Fe K$\alpha$ line fluxes, a
nickel-to-iron abundance ratio was estimated at 0.055--0.075, a
factor of 1.5--2 larger than the cosmic values \citep{And89}.
Similar studies of the Seyfert~2 galaxy NGC 1068 were reported by
\cite{Mat04} using {\em Chandra} and {\em XMM-Newton}
observations. Fe and Ni fluorescence emission in both neutral and
highly ionized material was identified and, from the
nickel-to-iron flux ratio, Fe was found to be overabundant by a
factor of 2 with respect to solar and Ni by the same factor with
respect to iron. \cite{Pou06} confirmed the nickel emission in NGC
1068 as well as in Mkn 3, a Seyfert~2 galaxy previously studied by
\cite{Pou05}. \cite{Pou06} also showed evidence of high-velocity
shifts in the lines coming from ionized nickel. \cite{Mar07}
detected with {\em Suzaku} narrow K$\alpha$ fluorescence emission
lines from Fe, Si, S, Ar, Ca, and Ni in the radio-loud AGN
Centaurus A (NGC 5128).

\cite{Mol98} analyzed data collected with {\em BeppoSAX} from the
Perseus cluster of galaxies, to find that the ratio of the flux of
the 7--8 keV line complex to the 6.7 keV line was significantly
larger than that predicted by optically thin plasma codes, and
that it diminishes with increasing cluster radius. It is argued
that this effect is due to resonance scattering in an optically
thick plasma at the energies of the Fe K$\alpha$ line. However,
\cite{Gas04} used {\em XMM-Newton} observations of the same
cluster to measure an overabundance for nickel. These authors
claimed that the excess in the flux of the 7--8 keV line complex
was due to Ni K$\alpha$ emission rather than resonance scattering.
\cite{deP04} also detected an overabundant nickel in the {\em
XMM-Newton} spectrum of the cluster Abel 478, but they noticed
that this could be overestimated due to errors in the nickel line
energies. The determination of the Ni abundance profile in the ICM
is also important because Ni is almost exclusively produced in
SNIa; therefore, high resolution spectral analysis, together with
accurate atomic data, are needed to understand these observations.

The first object detected by {\em INTEGRAL} in the Galactic plane,
IGR J16318-4848, is a possible X-ray binary where the obscuring
matter has a column density as large as the inverse Thompson cross
section. This source also shows strong Fe K$\alpha$ and Fe
K$\beta$ emission accompanied by a weaker, but nevertheless
distinct, Ni K$\alpha$ feature at 7.5 keV \citep{Wal03}.
Observations of the same source with {\em XMM-Newton} reveal
similar properties \citep{Mat03}, and a subsequent monitoring
campaign with both {\em INTEGRAL} and {\em XMM-Newton} confirmed
column densities of $1.2 \times 10^{24}$ cm$^{-2}$ and large
equivalent widths for the K$\alpha$ emission lines of Fe and Ni
\citep{Iba07}.

Perhaps the first unambiguous detection of Ni K lines in an
astrophysical spectrum is due to \cite{koy07}. They have observed
the diffuse X-ray emission from the Galactic center using the
X-ray Imaging Spectrometer (XIS) on {\em Suzaku}. They have
detected, for the first time, lowly ionized nickel and He-like
nickel K lines, and have measured the K$\alpha$ line flux ratio of
Ni~{\sc xxvii} to that of Fe~{\sc xxv} to determine a plasma
temperature of $\sim$5.4 keV, assuming solar relative Ni and Fe
abundances.

Following work by \cite{pal02,pal03a,pal03b}, \cite{bau03,bau04},
\cite{men04}, and \cite{kal04} on the Fe K lines, by \cite{gar05}
on the K-shell photoabsorption of O ions, and the recent study of
the K lines in the Ne, Mg, Si, Ar, and Ca isonuclear sequences
\citep{pal08}, we report new atomic data for K-vacancy levels in
the nickel isonuclear sequence. Prime objectives are to improve
the atomic database of the XSTAR modelling code \citep{bau01} and
to prepare ionic targets (configuration expansions and orbitals)
for the lengthy computations of the K-shell photoabsorption and
photoionization cross sections, where both radiative and Auger
dampings are key effects. With respect to Ni, available atomic
structure data sets---namely K-vacancy level energies,
wavelengths, $A$-values, and radiative and Auger widths---for
first-row ions with electron number $2 \le N \le 9$ are far from
complete while for the second and third-row ions ($10 \le N \le
27$) they are certainly lacking. The only exception is Ni$^+$
where measurements in the solid state have been reported
\citep{sal70, sal72, sli72, ber78, rao86, per87, hol97, raj01,
egr08} and the K$\alpha $ unresolved transition array (UTA)
centroid wavelength have been calculated by \cite{hou69}.

Previous study on the K-shell structure of nickel includes that by
\cite{hsu87} on the satellite spectra of He-like nickel. They have
recorded spectra emitted from the plasma of the Tokamak Fusion
Test Reactor (TFTR) with a high-resolution crystal spectrometer,
providing a K-line list for Ni~{\sc xxvii} ($N=2$) and  Ni~{\sc
xxvi} ($N=3$) and interpreting the observed spectra with the aid
of a Hartree-Fock-Slater (HFS) calculation. \cite{bom88} have
measured the wavelengths of the K lines in Ni~{\sc xxiv}--Ni~{\sc
xxvii} ($N=$ 2--5) emitted by a hot plasma at the Joint European
Torus (JET). In order to analyze these observations, they have
computed wavelengths, $A$-values, and radiative widths with the
SUPERSTRUCTURE atomic structure code \citep{eis74} and Auger rates
with the AUTOLSJ \citep{tfr81}. \cite{vai78} have calculated
wavelengths, radiative transition probabilities, and
autoionization (Auger) rates for ions with atomic numbers $Z=$
4--34 using the $1/Z$ expansion technique. They have considered
the ${\rm 1s-2p}$ transitions in the H-like sequence, ${\rm
1s2l-2p2l}$ and ${\rm 1s^2-1s2l}$ in the He-like sequence, and
${\rm 1s^22l-1s2p2l}$ in the Li-like sequence. Energies for
K-vacancy levels of the type ${\rm 1s}nl$ in the He-like
isoelectronic sequence have been calculated by \cite{vai85} using
the same technique. \cite{gor03} have audited the fluorescence
database by \cite{kaa93}, which is widely used in modelling codes,
in particular their scaling procedures along isoelectronic
sequences. They have found serious flaws which appear to
compromise the application of this database in plasma modelling.

The outline of the present report is as follows. The numerical methods are
briefly decribed in Section~2 while an analysis of the results based on
comparisons with previous experimental and theoretical values is carried
out in Section~3. The two supplementary electronic tables are explained
in Section~4 while some conclusions are finally discussed in Section~5.

\section{Numerical methods}

Three independent atomic structure packages have been used. The
main body of atomic data is computed with HFR, the Hartree--Fock
with Relativistic corrections method of \cite{cow81}. Data
accuracy is assessed by means of two other approaches: the
multiconfiguration Breit--Pauli method, which incorporates a
scaled Thomas--Fermi--Dirac statistical potential as implemented
in AUTOSTRUCTURE \citep{eis74, bad86, bad97}, and the GRASP code
\citep{gra80a, gra80b, mck80} based on the multiconfiguration
Dirac--Fock method.

 In HFR and AUTOSTRUCTURE, wavefunctions are calculated with the
Breit--Pauli relativistic corrections
 \begin{equation}
H_{\rm BP} = H_{\rm NR} + H_{\rm 1B} + H_{\rm 2B}
\end{equation}
where $H_{\rm NR}$ is the usual non-relativistic Hamiltonian. The
one-body relativistic operators
\begin{equation}
H_{\rm 1B} = \sum^{N}_{n=1} f_n(\rm mass) + f_n(\rm d) + f_n (\rm so)
\end{equation}
represent the spin--orbit interaction, $f_n(\rm so)$, the
non-fine-structure mass variation, $f_n(\rm mass)$, and one-body
Darwin correction, $f_n(\rm d)$. The two-body Breit operators are
given by
\begin{equation}
H_{\rm 2B} = \sum_{n<m} g_{nm}(\rm so) + g_{nm}(\rm ss) + g_{nm}
(\rm css) + g_{nm}(\rm d) + g_{nm}(\rm oo) \label{breit}
\end{equation}
where the fine-structure terms are $g_{nm}(\rm so)$
(spin-other-orbit and mutual spin-orbit) and $g_{nm}(\rm ss)$
(spin-spin), and the non-fine-structure counterparts are
$g_{nm}(\rm css)$ (spin-spin contact), $g_{nm}(\rm d)$ (two-body
Darwin), and $g_{nm}(\rm oo)$ (orbit-orbit). HFR computes
energies, $A$-values, and Auger rates with non-orthogonal orbital
bases, which are generated by optimizing the average energy of
each configuration. It also neglects the part of the Breit
interaction (Eq.~(\ref{breit})) that cannot be reduced to a
one-body operator. AUTOSTRUCTURE can use both orthogonal and
non-orthogonal orbital bases for all the electronic configurations
considered which enables estimates of relaxation effects. We have
used the same configuration-interaction (CI) expansions as in our
previous papers on the Fe isonuclear sequence \citep{bau03,
pal03a, pal03b, men04}.

Auger rates are computed in both HFR and AUTOSTRUCTURE in a
distorted wave approach. However, as in the Fe isonuclear sequence
\citep{pal03b}, the level-to-level computation of Auger rates in
ions with open 3d shell ($N > 17$) proved to be intractable;
therefore, we have employed the formula given in \cite{pal01} for
the single-configuration average (SCA) Auger decay rate.

Our third package is GRASP where the atomic state function (ASF)
is represented as a superposition of configuration state functions
(CSF) of the type
\begin{equation}
\Psi(\alpha \Pi JM) = \sum_{i} c_i(\alpha)\Phi(\beta_i \Pi JM)
\label{mcdf}
\end{equation}
where $\Psi$ and $\Phi$ are respectively the ASF and CSF. $\Pi$,
$J$, and $M$ are the relevant quantum numbers, i.e., parity, total
angular momentum, and its associated total magnetic number,
respectively. $\alpha$ and $\beta_i$ stand for all the other
quantum numbers that are necessary to describe unambiguously the
ASFs and CSFs. The summation in Eq.~(\ref{mcdf}) is up to $n_c$,
the number of CSFs in the expansion, and each CSF is built from
antisymmetrized products of relativistic spin orbitals. The $c_i$
coefficients, together with the orbitals, are optimized by
minimizing an energy functional. The latter is built from one or
more eigenvalues of the Dirac-Coulomb Hamiltonian depending upon
the optimization option adopted. Here, we have used the extended
average level (EAL) option in which the $(2J+1)$-weighted trace of
the Hamiltonian is minimized. Transverse Breit interaction as well
as other QED interactions, e.g. the vacuum polarization and
self-energy, have been included in the Hamiltonian matrix as
perturbations. This code does not treat the continuum, and has
thus been exclusively employed in comparisons of radiative data
for bound-bound transitions.

%% In a manner similar to \objectname authors can provide links to dataset
%% hosted at participating data centers via the \dataset{} command.  The
%% second curly bracket argument is printed in the text while the first
%% parentheses argument serves as the valid data set identifier.  Large
%% lists of data set are best provided in a table (see Table 3 for an example).
%% Valid data set identifiers should be obtained from the data center that
%% is currently hosting the data.
%%
%% Note that AASTeX interprets everything between the curly braces in the
%% macro as regular text, so any special characters, e.g. "#" or "_," must be
%% preceded by a backslash. Otherwise, you will get a LaTeX error when you
%% compile your manuscript.  Special characters do not
%% need to be escaped in the optional, square-bracket argument.

%% In this section, we use  the \subsection command to set off
%% a subsection.  \footnote is used to insert a footnote to the text.


%% This section also includes several of the displayed math environments
%% mentioned in the Author Guide.

\section{Results}

Detailed comparisons with previous data have been carried out in
order to obtain accuracy estimates and detect weak points. In the
following sections, we give a concise account of our HFR
computations of level energies, radiative and Auger widths of
K-vacancy states, and wavelengths and radiative transition
probabilities for K lines. With respect to the K$\alpha_{1,2}$ and
K$\beta$ unresolved transition arrays (UTAs) characteristic in
second- and third-row ions ($12\leq N\leq 27$), we have determined
centroid wavelengths and $K\alpha_2/K\alpha_1$ and
$K\beta/K\alpha$ line ratios which can be compared to the
available solid-state measurements \citep{sal70, sal72, sli72,
ber78, rao86, per87, hol97, raj01}. The $KLM/KLL$ and $KMM/KLL$
Auger decay channel ratios are also investigated along the
isonuclear sequence, and HFR relative Auger channel intensities in
Ni$^+$ are compared with a recent experiment \citep{egr08}.
Finally, the trends of the K-shell fluorescence yield and natural
K-level width with electron number $N$ are studied.

\subsection{Energy levels}

K-vacancy level energies for Ni ions are very scarce in the
literature. The NIST database \citep{nist} lists values for He-
and Li-like Ni that are interpolated or extrapolated from
experimental level energies along isoelectronic sequences.
\cite{vai85} calculated values in Ni$^{26+}$ using the $1/Z$
expansion technique. In Table~\ref{tablev}, we compare our HFR
level energies with the two above-mentioned data sets, finding an
agreement within 3 eV.

\subsection{Wavelengths}

The only wavelength measurements for Ni K lines available in the
literature are those obtained for highly charged ions ($N=$ 2--5)
in tokamak plasmas \citep{hsu87,bom88} and in the solid state with
the recent experiment of \cite{hol97}. Previous calculations are
due to: \cite{vai78,vai85} using the $1/Z$ expansion technique for
He- and Li-like Ni; \cite{hsu87} by means of the
Hartree--Fock--Slater (HFS) method to interpret observations in
He- and Li-like Ni; and \cite{bom88} with SUPERSTRUCTURE to model
measurements in He- to B-like Ni.

These different sets of wavelengths are compared to our HFR
calculation in Table~\ref{tabwl} for ions with $N=$2--5. The
agreement with experiment, the $1/Z$ expansion, and the HFS
calculations is within 1~m\AA. Concerning the somewhat larger
discrepancies encountered with the SUPERSTRUCTURE calculation of
\cite{bom88}, we believe that they are due to their optimization
procedure since they optimized the atomic orbitals in the Li-like
system and kept them fixed for all the other ions.

\begin{figure}
\epsscale{1.} \plotone{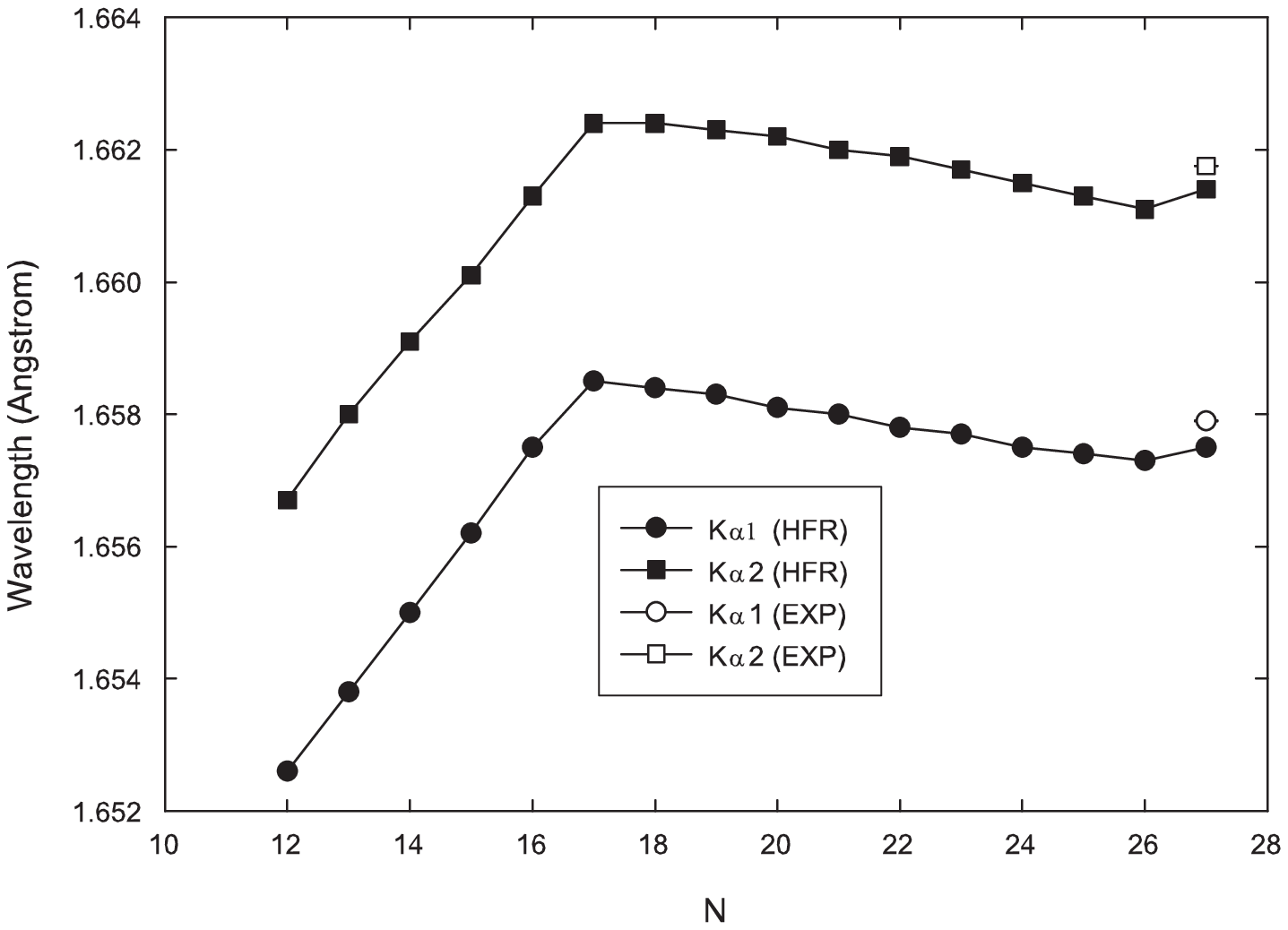}
\caption{HFR wavelengths (\AA) for the $K\alpha1$ (filled circles) and
$K\alpha2$ (filled squares) UTAs in Ni ions with electron number
$12\leq N\leq 27$. The solid-state measurements by \cite{hol97} are
also shown: open circle, $K\alpha1$; open square, $K\alpha2$. \label{fig1}}
\end{figure}

\begin{figure}
\plotone{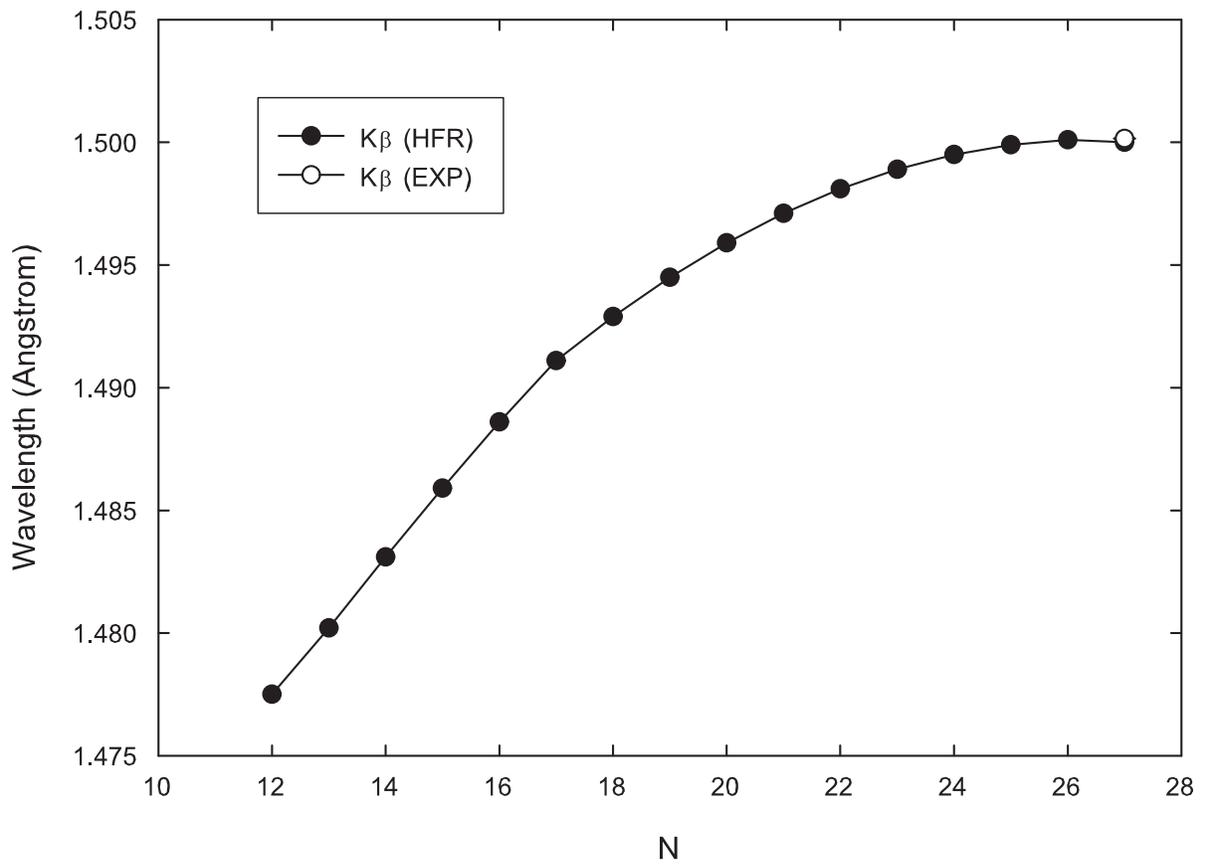}
\caption{HFR wavelengths (\AA) of the $K\beta$ UTA for Ni ions with electron
number $12\leq N\leq 27$ (filled circles). The solid-state measurement
by \cite{hol97} is also shown (open circle).\label{fig2}}
\end{figure}

HFR centroid wavelengths in the second- and third-row ions ($N=$
12--27) are plotted as a function of $N$ for the K$\alpha_{1,2}$
(Figure~\ref{fig1}) and K$\beta$ (Figure~\ref{fig2}) UTAs. It may
be seen that the K$\alpha_{1,2}$ lines are sharply shifted to the
red as $N$ increases up to $N=17$, and from then on slowly shifted
to the blue. This was also the case in our previous studies of the
Fe K lines \citep{pal03a,pal03b}, and is consistent with the trend
calculated by \cite{hou69}. The K$\beta$ line, in contrast, is
monotonically shifted to the red with an increasing $N$. The
values measured by \cite{hol97} for the K$\alpha_1$, K$\alpha_2$,
and K$\beta$ solid-state UTAs are also shown in both figures,
namely $\lambda\lambda$ 1.65790(1), 1.66175(1), and 1.500152(3),
respectively. They compare favorably with our HFR values for
Ni$^+$ ($N=27$): $\lambda\lambda$ 1.6575, 1.6614, 1.5000. As for
the highly charged ions, the accord is better than 1~m\AA.

\subsection{$A$-values, radiative widths and line ratios}

\begin{figure}
\epsscale{0.8}
\plotone{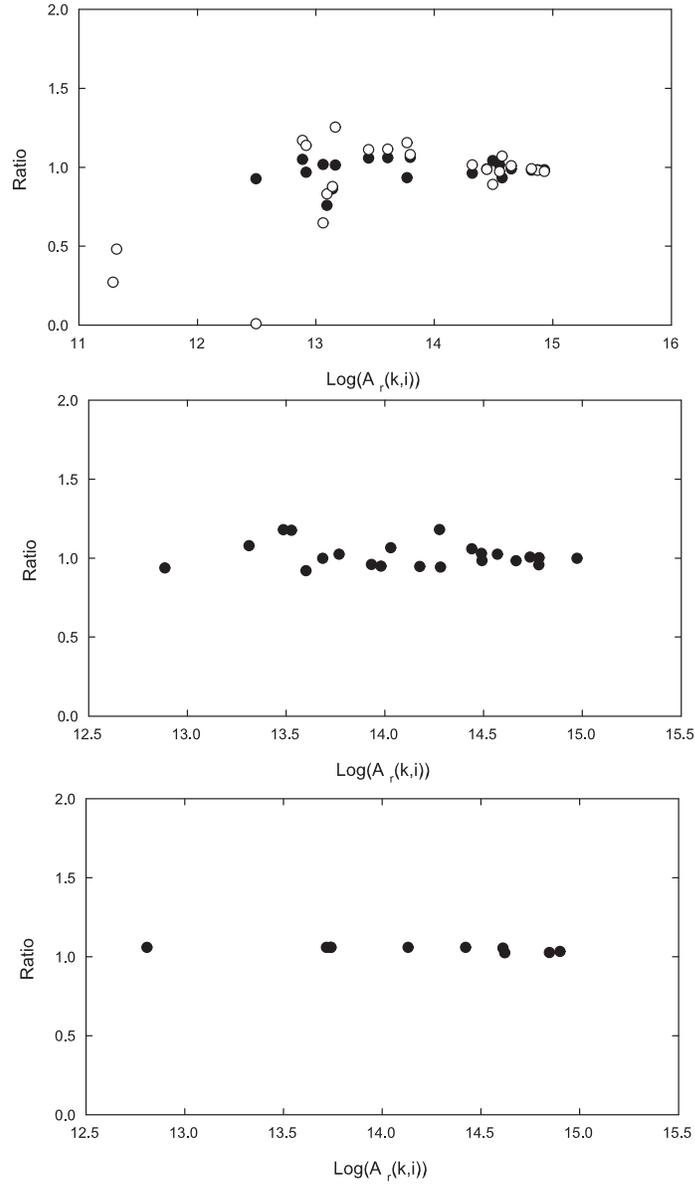}
\caption{Comparison of the present HFR $A$-values for K transitions
in Li-like (top panel), Be-like (middle panel) and B-like nickel (bottom panel)
with two independent calculations. Filled circles: \cite{bom88}. Open circles:
\cite{vai78}.
\label{fig3}}
\end{figure}

A comparison of our HFR $A$-values for K transitions in ions with
$3\leq N\leq 5$ with the data of \cite{bom88} and \cite{vai78} is
given in Figure~\ref{fig3}, where the ratios with respect to HFR
are plotted as a function of the HFR $A$-value. A large scatter is
observed for $A\lesssim 10^{13}$ s$^{-1}$, especially with the
$A$-values for the Li-like system by \cite{vai78}, thus
illustrating the model dependency of the weak rates. The average
ratios in the Li-like species are 0.98$\pm$0.07 for \cite{bom88}
and 0.91$\pm$0.31 for \cite{vai78}. The 31\% standard deviation
observed in the latter is due to the weaker rates which are not
listed by \cite{bom88}. Concerning the Be- and B-like ions, where
the only available data are given in \cite{bom88}, the average
ratios are respectively 1.02$\pm$0.08 and 1.04$\pm$0.01. Based on
these comparisons, we are confident that our $A$-values have a
10\% accuracy for transitions with $A\gtrsim 10^{13}$ s$^{-1}$.

\begin{figure}
\epsscale{0.8}
\plotone{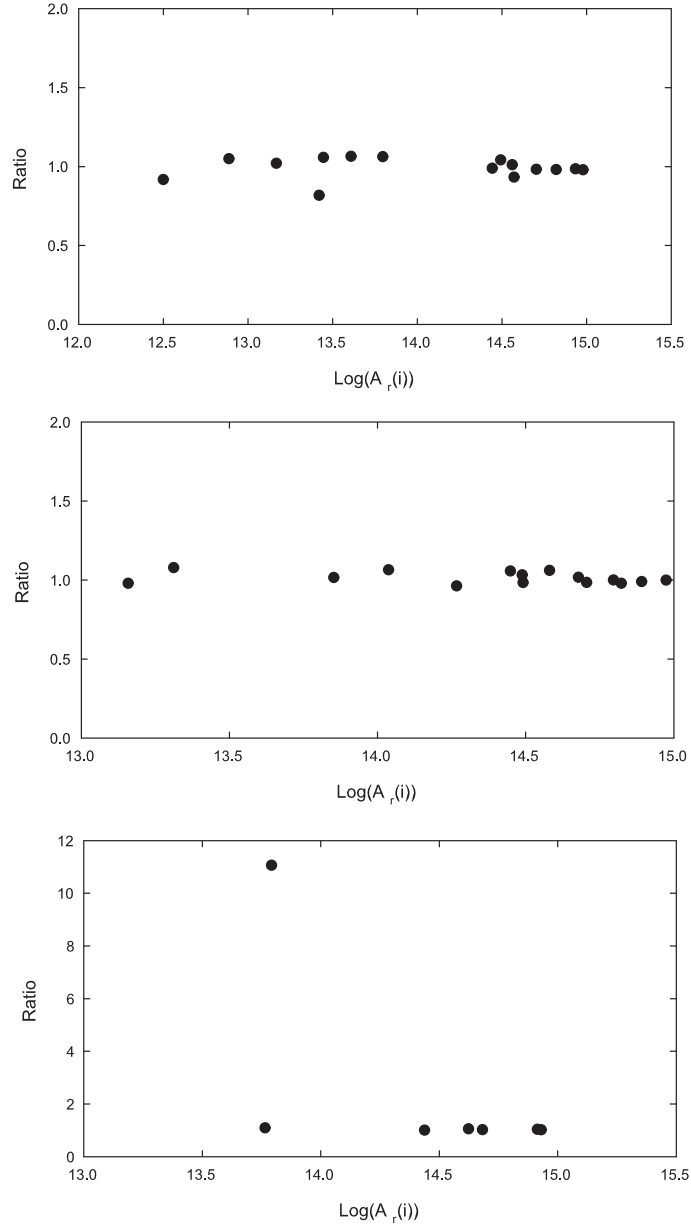}
\caption{Comparison of the present HFR radiative widths for K-vacancy levels
in Li-like (top panel), Be-like (middle panel) and B-like nickel (bottom panel)
with those by \cite{bom88}.
\label{fig4}}
\end{figure}

In Figure~\ref{fig4}, we compare our HFR radiative widths with the
values obtained by \cite{bom88} for systems with $3\leq N\leq 5$.
The ratio with respect to HFR is plotted as a function of the HFR
width. If the ${\rm 1s2s}^2{\rm 2p}^2\,^4{\rm P}_{1/2}$ level in
the B-like ion is excluded from this comparison, the average
ratios are then 0.99$\pm$0.07 for $N=3$, 1.01$\pm$0.04 for $N=4$,
and 1.04$\pm$0.03 for $N=5$. Concerning the excluded level, the
width of 6.86$\times$10$^{14}$ s$^{-1}$ quoted by \cite{bom88} is
questionable. It differs by one order of magnitude with our HFR
width (6.20$\times$10$^{13}$ s$^{-1}$) and with the value they
list for the other level of the same multiplet, namely
6.35$\times$10$^{13}$ s$^{-1}$ for ${\rm 1s2s}^2{\rm 2p}^2\,^4{\rm
P}_{5/2}$. Thus, this error must be due to a misprint and the
agreement is better than 10\%.

\begin{figure}
\plotone{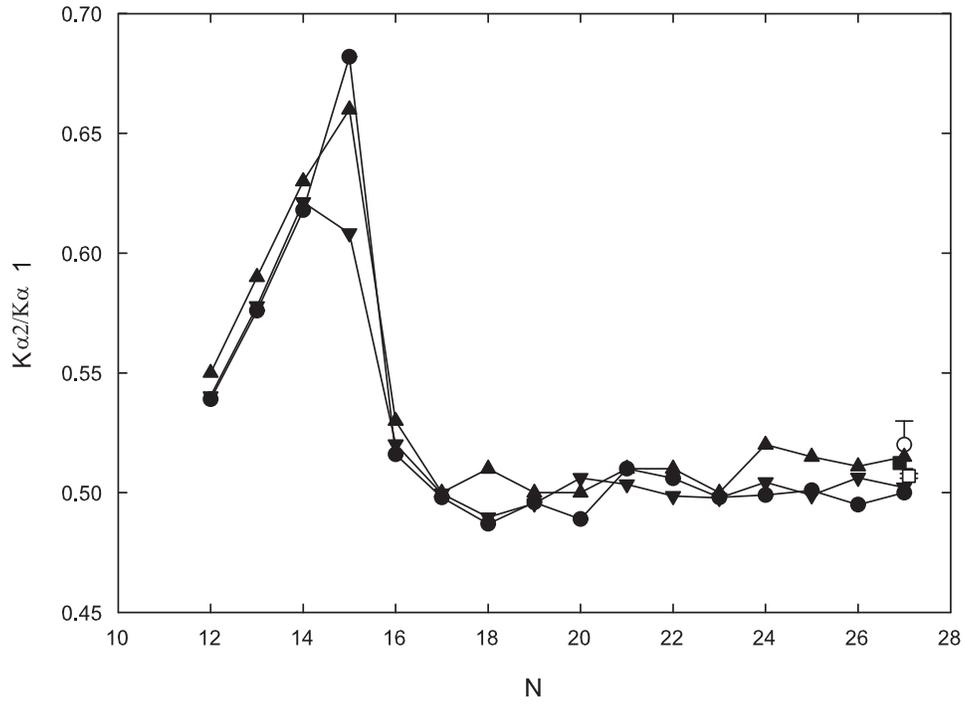}
\caption{$K\alpha_2/K\alpha_1$ line ratio for Ni ions with electron number
$12\leq N\leq 27$. Filled circle: HFR (this work). Filled upright triangles:
MCDF-EAL (this work). Filled inverted triangles: AUTOSTRUCTURE (this work).
Filled squares: Dirac--Fock calculation \citep{sco74}.
Open circle: solid-state measurement by \cite{hol97}. Open square:
measurement in the solid by \cite{sal70}.\label{fig5}}
\end{figure}

The trend of the $K\alpha_2/K\alpha_1$ line ratio along the
isonuclear sequence is shown in Figure~\ref{fig5} for the
K$\alpha_{1,2}$ UTAs in members with $12\leq N\leq 27$. This line
ratio increases with $N$ up to $N=15$, and then decreases to
become almost constant at a value of $\sim$0.5 for $N\ge 17$. This
trend is consistent (within 10\%) in the three numerical
approaches that we considered, i.e. HFR, AUTOSTRUCTURE, and MCDF.
Moreover, our calculated line ratios for $N=27$ are compared with
the Dirac--Fock value of \cite{sco74} and the two solid-state
measurements of \cite{sal70} and \cite{hol97}. All these results
agree to within 5\%.

\begin{figure}
\epsscale{0.8}
\plotone{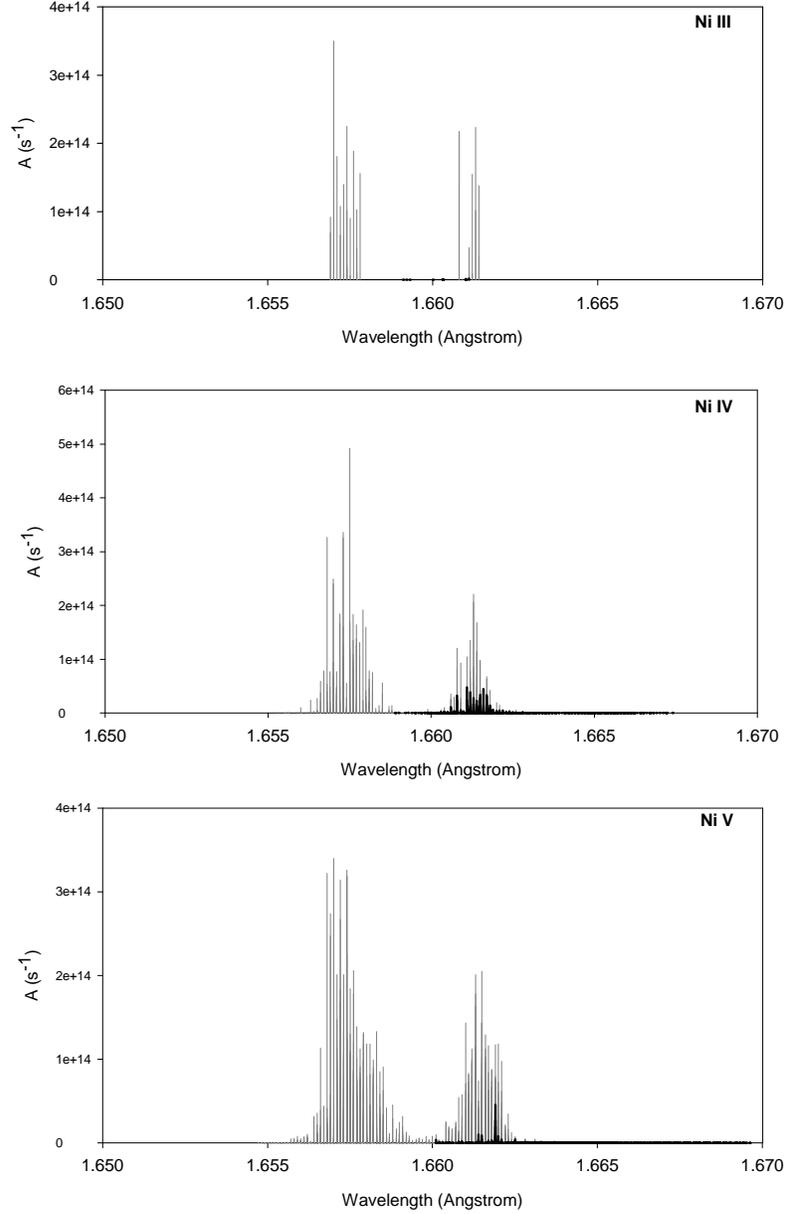}
\caption{Stick spectra, i.e. $A$-value (s$^{-1}$) vs. wavelength (\AA),
in the region of the Ni $K\alpha$ line computed with HFR
in Ni~{\sc iii} (top panel),  Ni~{\sc iv} (middle panel), and Ni~{\sc v}
(bottom panel). Thin gray sticks are diagram lines, ${\rm [1s]3d}^M - {\rm [2p]3d}^M$,
while the thick black sticks are the ${\rm [1s]3d}^M - {\rm [2p]3d}^{M-1}{\rm 4s}$ and
${\rm [1s]3d}^M - {\rm [2p]3d}^{M-2}{\rm 4s^2}$ satellite lines. \label{fig6}}
\end{figure}

The contribution of satellites to the $K\alpha_2/K\alpha_1$ line
ratio in Ni$^{3+}$ is illustrated in Figure~\ref{fig6}. Stick
spectra, i.e. $A$-values as function of the transition wavelength,
computed with HFR are plotted in the region of the K$\alpha$ line
for the diagram lines (${\rm[1s]3d}^M - {\rm [2p]3d}^M$) in thin
gray sticks and for the satellites (${\rm [1s]3d}^M - {\rm
[2p]3d}^{M-1}{\rm 4s}$ and ${\rm [1s]3d}^M -{\rm [2p]3d}^{M-2}{\rm
4s^2}$) in thick black sticks. The top panel shows the Ni$^{2+}$
spectrum, the middle Ni$^{3+}$, and the bottom Ni$^{4+}$. One
clearly sees that the satellites in Ni$^{3+}$ are intense relative
to the K$\alpha_2$ diagram lines and are blended with the latter.
As a consequence, the $K\alpha_2/K\alpha_1$ line ratio drops to
0.37 without satellites instead of having a value of 0.50 when
included. In the other ions, their effect on the line ratio can be
neglected.

\begin{figure}
\plotone{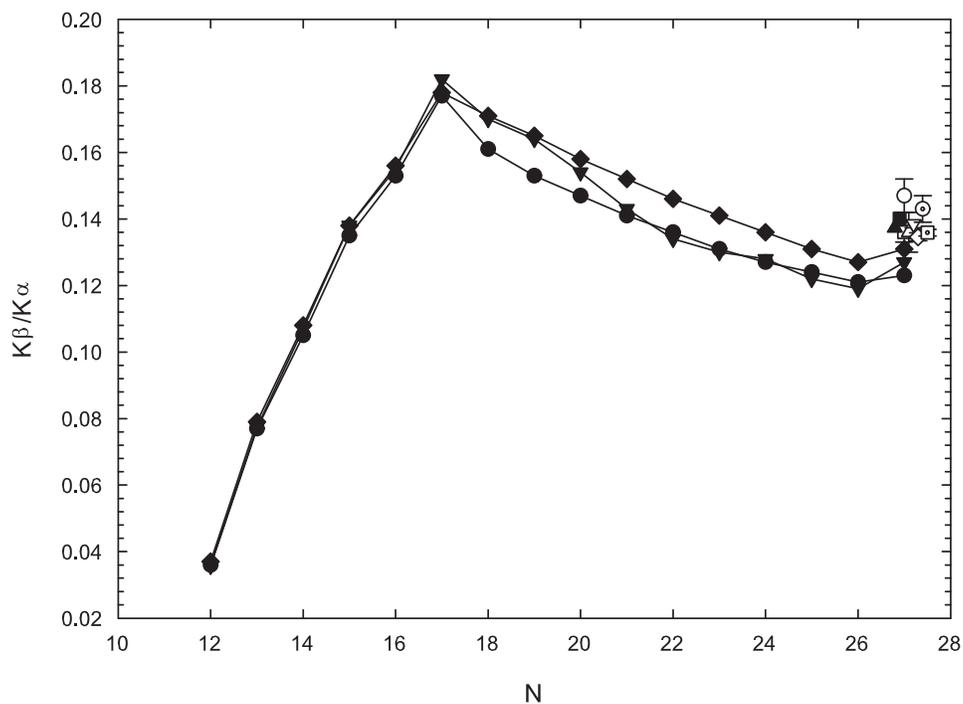}
\caption{$K\beta/K\alpha$ line ratio for Ni ions with electron number $12\leq N\leq 27$.
Filled circles: HFR (this work). Filled inverted triangles: AUTOSTRUCTURE (this work).
Filled diamonds: MCDF-EAL (this work). Filled square: Dirac--Fock calculation
\citep{sco74}. Filled upright triangle: MCDF-SAL calculation \citep{pol98}.
The solid-state measurements are as follows. Open circle:
\cite{hol97}. Open square: \cite{sal72}. Open upright triangle: \cite{rao86}.
Open inverted triangle: \cite{sli72}.
Open diamond: \cite{raj01}.
Dotted circle: \cite{ber78}.
Dotted square: \cite{per87}.\label{fig7}}
\end{figure}

The $K\beta/K\alpha$ line ratio is plotted as function of $N$ in
Figure~\ref{fig7} for the K$\alpha$ and K$\beta$ UTAs in ions with
$12\leq N\leq 27$. One can see that the trend along the isonuclear
sequence presents a sharp increase with $N$ up to $N=17$ due to
the filling of the 3p subshell, the fertile domain of the K$\beta$
transition, followed by a slow decrease associated with the
filling of the 3d subshell. This behavior is confirmed with our
three independent numerical methods (HFR, AUTOSTRUCTURE, and MCDF)
to within 10\%.

In Figure~\ref{fig7}, our computed $K\beta/K\alpha$ line ratios
for $N=27$ are compared with the solid-state measurements
\citep{sal72, sli72, ber78, rao86, per87, hol97, raj01} and other
theoretical values \citep{sco74,pol98}. The bulk of the
experimental values are located at 0.139$\pm$0.015 (this deviation
takes into account the experimental error bars) which can be
compared to the ratios we computed with HFR (0.117), AUTOSTRUCTURE
(0.126), and MCDF (0.131). Other theoretical values are also close
to experiment, namely the Dirac-Fock result by \citet{sco74} at
0.140 and that with the MCDF-SAL method of 0.137 \citep{pol98}.
Therefore, there is good theory--experiment agreement in Ni$^+$
except for our HFR ratio which is 15\% lower.

\subsection{Auger widths and Auger channel ratios}

\begin{figure}
\plotone{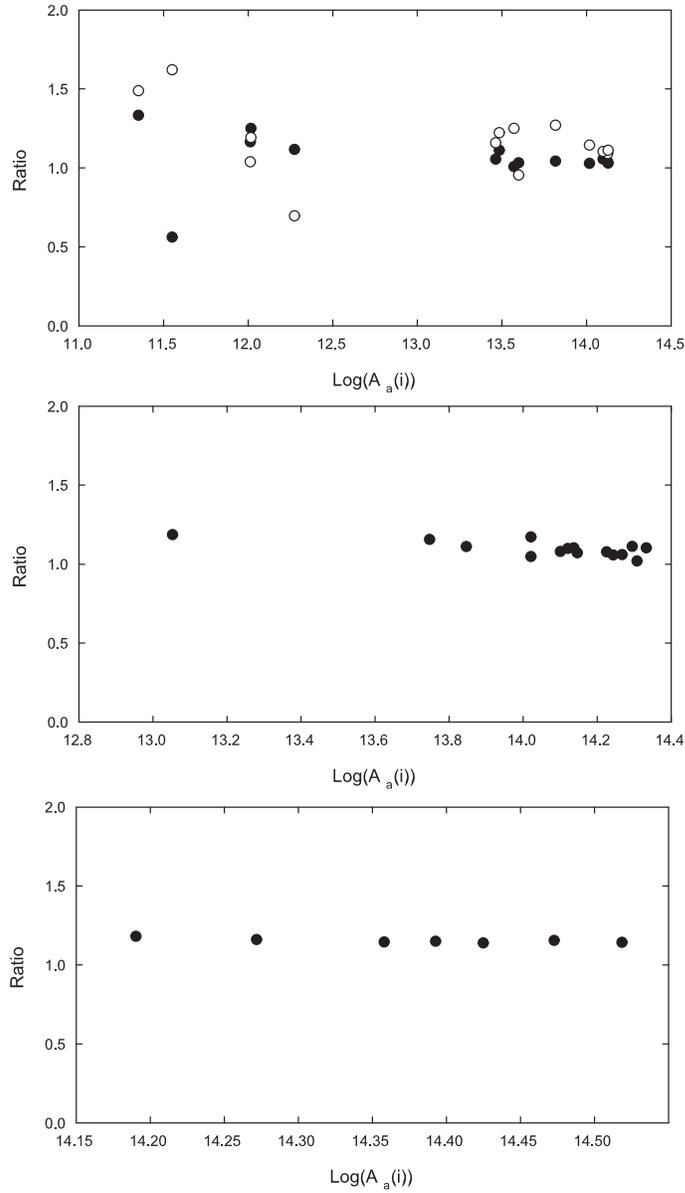}
\caption{Comparison of the present HFR Auger widths for K-vacancy levels
in Li-like (top panel), Be-like (middle panel) and B-like nickel (bottom panel)
with two independent calculations. Filled circles: \cite{bom88}. Open circles:
\cite{vai78}.
\label{fig8}}
\end{figure}

For ions with electron number $3\leq N\leq 5$, we compare HFR
Auger widths in Figure~\ref{fig8} with the results by \cite{bom88}
and \cite{vai78}. Ratios with respect to HFR are plotted as a
function of the HFR Auger width. Large scatter may be observed for
$A_{\rm a}\lesssim 10^{13}$ s$^{-1}$. Excluded from this figure
are the ratios for the ${\rm1s(^2S)2s2p(^3P^o)\,^2P^o_{3/2}}$
level in the Li-like system which has a small width, namely  $A_a
(HFR) = 6\times10^9$ s$^{-1}$, for which large discrepancies are
found. If ratios less than 10$^{13}$ s$^{-1}$ are discarded, the
average ratios in Ni$^{25+}$ are 1.04$\pm$0.03 \citep{bom88} and
1.14$\pm$0.10 \citep{vai78}. The accord between these three
independent methods is fair, although the widths obtained by the
latter are significantly higher ($\sim$15\% on average) than HFR.
Concerning the other two ions, where the only available data are
those by \cite{bom88}, the average ratios are 1.10$\pm$0.05
(Ni$^{24+}$) and 1.15$\pm$0.03 (Ni$^{23+}$). These systematic
deviations (10\% and 15\% higher than HFR, respectively in the Be-
and B-like species) may be due, as previously mentioned, to the
fact that \cite{bom88} used orbitals optimized in the Li-like
system.

\begin{figure}
\plotone{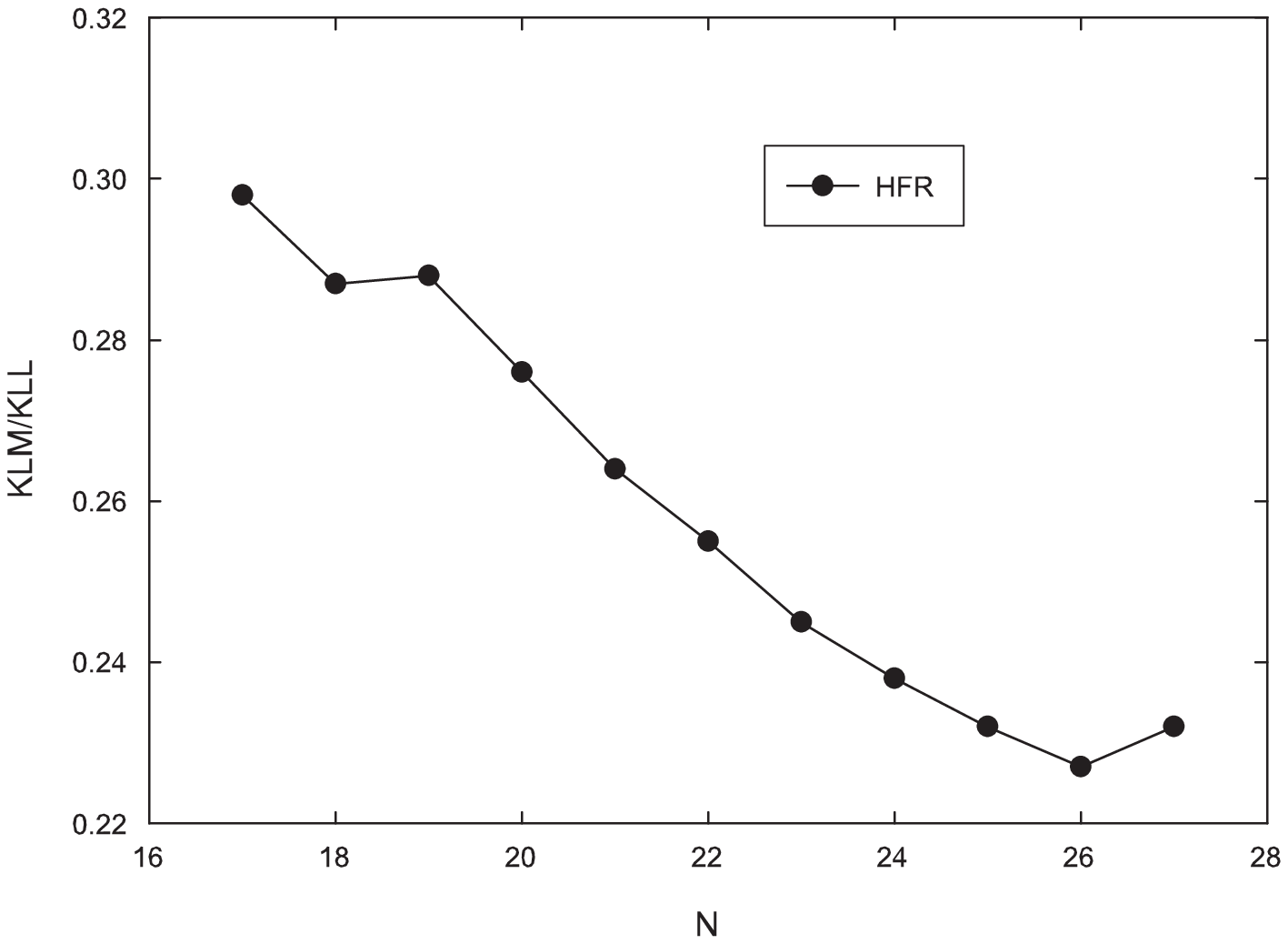}
\caption{HFR $KLM/KLL$ Auger channel ratio for Ni ions with electron number
$17\leq N\leq 27$.\label{fig9}}
\end{figure}

\begin{figure}
\plotone{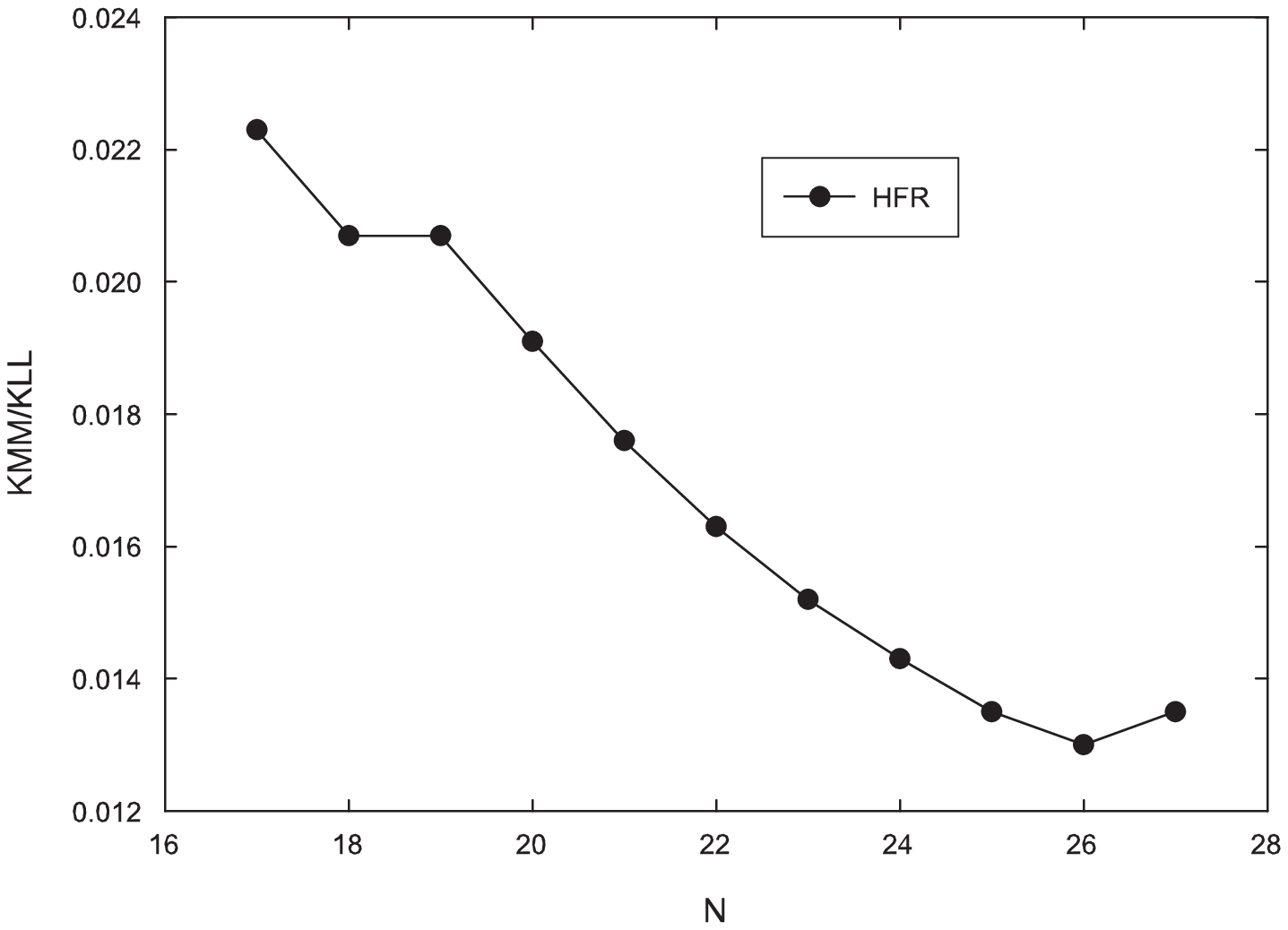}
\caption{HFR $KMM/KLL$ Auger channel ratio for Ni ions with electron number
$17\leq N\leq 27$. \label{fig10}}
\end{figure}

The trends of the HFR $KLM/KLL$ and $KMM/KLL$ channel ratios along
the isonuclear sequence ($17\leq N\leq 27$) are depicted in
Figures \ref{fig9} and \ref{fig10}, respectively, and are found to
be similar to those previously observed in iron \citep{pal03b}.
Unfortunately, here there is no available solid-state measurements
to compare with the theoretical ratios.

In Table~\ref{tabklm}, the HFR KLM Auger channel relative
intensities in Ni$^+$ are compared with the recent measurements of
\cite{egr08} made in the solid. Both data sets agree within the
experimental error ($\sim$20\%).

\subsection{Fluorescence yields and natural K-level widths}

\begin{figure}
\plotone{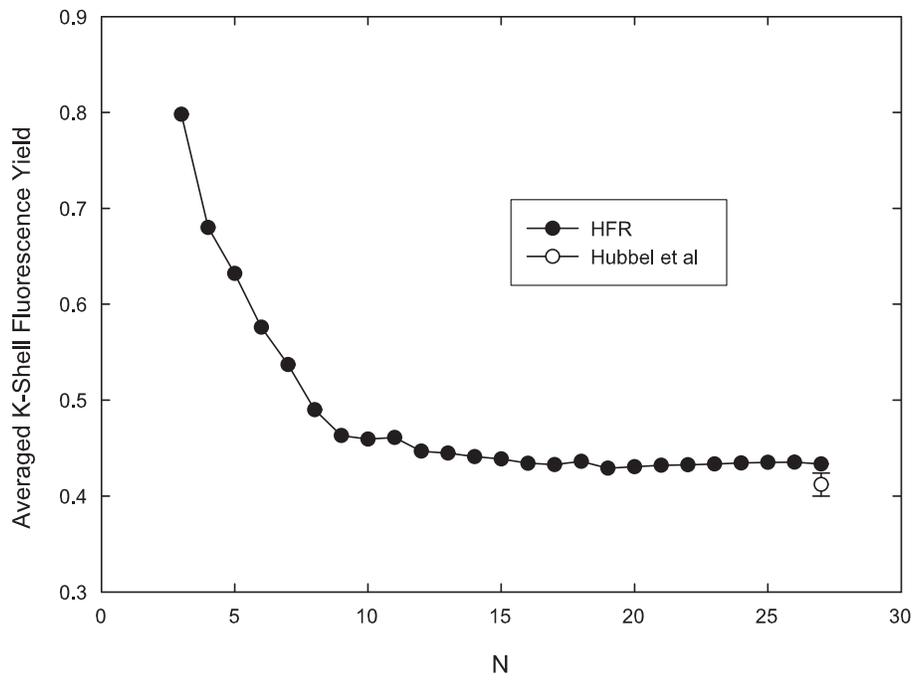}
\caption{HFR averaged K-shell fluorescence yield  for Ni ions with electron number
$3\leq N\leq 27$. The recommended value of \cite{hub94} is also shown. \label{fig11}}
\end{figure}

The HFR average K-shell fluorescence yield, $\overline{\omega}_K$,
for Ni ions with $3\leq N\leq 27$ is plotted in Figure~\ref{fig11}
as a function of $N$. For each ion, the average is computed over
all fine-structure levels
\begin{equation}
\overline{\omega}_K = \frac{1}{m} \sum_{i=1}^{m} \omega_K(i)
\end{equation}
where
\begin{equation}
\omega_K(i) = \frac{A_r(i)}{A_a(i)+A_r(i)}
\end{equation}
is the level yield. The following behavior may be noted:
$\overline{\omega}_K$ sharply decreases from 0.80 ($N=3$) to 0.46
($N=9$) in the first-row ions; then slowly decreases from 0.46
($N=10$) to 0.43 ($N=17$) in the second-row ions; and becomes
constant in the third-row ions at a value of 0.43 for Ni$^+$
($N=27$). The recommended value of \cite{hub94}, 0.412$\pm$0.012,
is also shown while our HFR value for $N=27$ is slightly higher by
less than 5\%.

\begin{figure}
\plotone{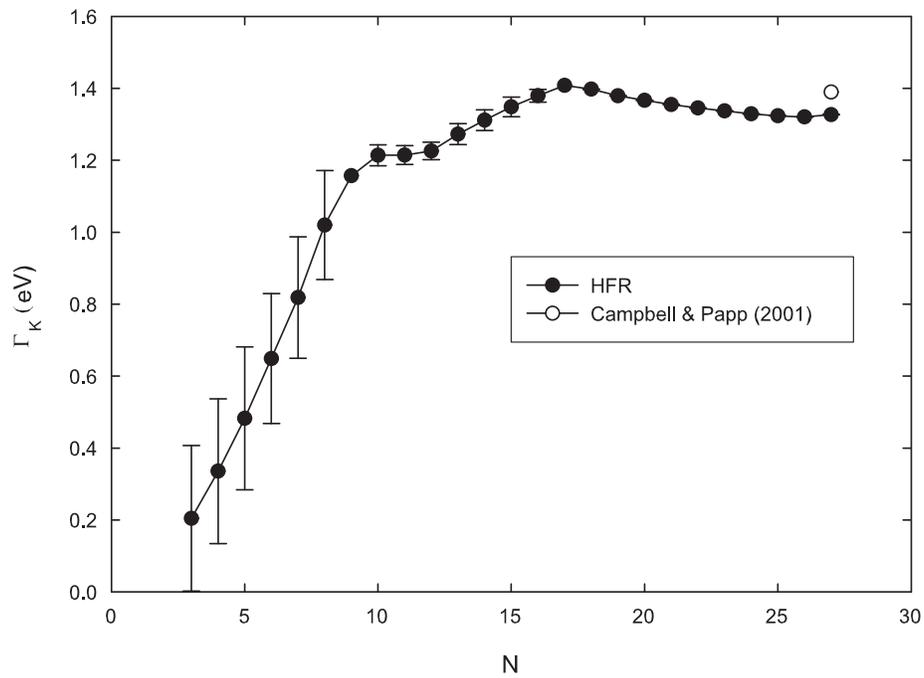}
\caption{HFR natural K-level width $\Gamma_K$ (eV) for Ni ions with electron number
$3\leq N\leq 27$. The average and standard deviation are plotted, the latter as error bars.
The recommended value of \cite{cam01} is also shown. \label{fig12}}
\end{figure}

In Figure~\ref{fig12}, the HFR natural K-level width, $\Gamma_K$
(in eV), is plotted along the nickel isonuclear sequence for ions
with $3\leq N\leq 27$. As for the fluorescence yields, the average
for each ion is given by
\begin{equation}
\overline{\Gamma}_K = \frac{1}{m} \sum_{i=1}^{m} \Gamma_K(i)
\end{equation}
with
\begin{equation}
\Gamma_K(i) = A_a(i)+A_r(i)
\end{equation}
and the standard deviation
\begin{equation}
\sigma(\Gamma_K) = \sqrt{\frac{1}{m}
\sum_{i=1}^{m}(\Gamma_K(i)-\overline{\Gamma}_K)^2}\ .
\end{equation}
$\overline{\Gamma}_K$ increases sharply from 0.20 eV for $N=3$ to
1.16 eV for $N=9$, delimiting the first-row regime; then, in the
second-row regime, it slowly increases from 1.21 eV for $N=10$ to
1.41 eV for $N=17$; and finally, it slowly decreases from 1.40 eV
for $N=18$ to 1.33 eV for $N=27$ in the third-row regime. This
last value agrees  within 5\% with the recommended natural K-level
width of 1.39 eV \citep{cam01} which is also plotted in the
figure. In the third-row regime, the natural K-level width of each
fine-structure level, $\Gamma_K(i)$, becomes constant in each ion
and thus $\sigma(\Gamma_K)$ is negligible as shown in the figure.

In Table~\ref{tabnatw}, the natural K-level widths recommended by
\cite{cam01} for the singly-ionized cosmically abundant elements
are compared with HFR values calculated in the present work and in
our earlier analyses \citep{pal03b, gar05, pal08}. The trend with
$Z$ is well reproduced and both data sets agree satisfactorily.

\section{Supplementary electronic tables}

Computed level energies, wavelengths, radiative transition
probabilities, absorption oscillator strengths, radiative and
Auger widths, and K-shell fluorescence yields in Ni$^+$-Ni$^{26+}$
can be accessed from the online ASCII Tables~\ref{klev}-\ref{klin}
through the VizieR Service at the CDS\footnote{\url
http://cdsweb.u-strasbg.fr}. The printed samples herein show data
for ions with electron number $N \le 3$.

It may be seen that in Table~\ref{klev} levels are identified with
the vector ($N$, $i$, $2S+1$,$L$, $2J$, ${\rm Conf}$) where $N$ is
the electron number, $i$ the level index, $2S+1$ the spin
multiplicity, $L$ the total orbital angular momentum quantum
number, $J$ the total angular momentum quantum number, and ${\rm
Conf}$ the level configuration assignment. For each level, the
computed HFR energy and its radiative width $A_r(i)$ are listed.
For K-vacancy levels, the Auger width $A_a(i)$ and the K-shell
fluorescence yield $\omega_K(i)$ are additionally given. In
Table~\ref{klin}, transitions are identified with the vector ($N$,
$k$,$i$) where $k$ and $i$ are the upper and lower level indices,
respectively, tabulating its computed wavelength $\lambda$,
radiative transition probabilities $A_r(k,i)$, weighted oscillator
strength $gf(i,k)$, and cancellation factor $CF$ as defined by
\cite{cow81}.

\section{Summary and conclusions}

Extensive data sets containing energy levels, wavelengths,
radiative transition probabilities, absorption oscillator
strengths, radiative and Auger widths, and fluorescence yields for
K-vacancy levels in the nickel isonuclear sequence have been
computed with the atomic structure codes HFR, AUTOSTRUCTURE, and
GRASP. For the ionic species Ni$^{2+}$-Ni$^{22+}$, with electron
numbers $6 \le N \le 26$, this is the first time that such data
become available. For ions with $2 \le N \le 5$ and $N=27$,
detailed comparisons have been carried out with available
measurements and theoretical values which bring forth the
consistency and accuracy of the present data.

Comparisons of HFR K-level energies with those reported in the
NIST database \citep{nist} for He- and Li-like nickel and the
calculated values in Ni$^{26+}$ by \cite{vai85} support an
accuracy rating for ions with electron number $N \le 3$ of better
than 3~eV. With regards to wavelengths, comparisons between HFR
and the tokamak data \citep{hsu87, bom88} for ions with $2\leq
N\leq 5$, with the solid-state measurements \citep{hol97} for
$N=27$, and with other published theoretical values \citep{vai78,
vai85, hsu87, bom88} result in a general agreement within 1~m\AA.

Based on the level of agreement of the present HFR $A$-values with
those by \citet{vai78} and \citet{bom88} for the highly ionized
species ($2\leq N\leq 5$), we are confident that they show an
accuracy within 10\% for transitions with $A\gtrsim 10^{13}$
s$^{-1}$. Regarding radiative widths, similar considerations have
led us to the conclusion that our HFR widths are also accurate to
10\%. For the $K\alpha_2/K\alpha_1$ and $K\beta/K\alpha$ line
ratios in the second and third-row ions ($12\leq N\leq 27$), the
trends along the isonuclear sequence are consistent (within 10\%)
for the three independent approaches we have used. For $N=27$, our
HFR $K\alpha_2/K\alpha_1$ ratio agrees within 5\% with the
solid-state measurements \citep{sal70, hol97}, but the HFR
$K\beta/K\alpha$ ratio is 15\% lower than the bulk of the
available experimental values located at 0.139$\pm$0.015.

Comparisons of the HFR Auger widths with those by \citet{bom88}
for species with $3\leq N\leq 5$ and by \cite{vai78} for $N=3$
show a fair consistency for widths greater than 10$^{13}$
s$^{-1}$, although the latter is higher by $\sim$15\% and the
former by 10\% for $N=4$ and 15\% for $N=5$. The systematic
deviations with the data by \cite{bom88} may be due to the orbital
optimization procedure they used. Moreover, recent measurements of
$KLM$ Auger channel relative intensities \citep{egr08} support our
HFR Auger rates. Finally, HFR fluorescence yield and natural
K-level width for $N=27$ are consistent within 5\% with the
recommended values found in the literature \citep{hub94,cam01}.

The present radiative and Auger widths will be used in the
computations of the K-shell photoionization cross sections of
these ions which are required in XSTAR \citep{kal01} for the
modelling of interesting Ni spectral features.

\acknowledgments

This work was funded in part by the NASA Astronomy and Physics Research
and Analysis Program. PP and PQ are Research Associates of the Belgian
FRS-FNRS.

\clearpage

\clearpage

\clearpage

\begin{deluxetable}{lllll}
\tabletypesize{\scriptsize}
%\rotate
\tablecaption{$K$-vacancy level energies (keV) for Ni ions with $2\leq N\leq 3$
\label{tablev}}
\tablewidth{0pt}
\tablehead{
\colhead{$N$} & \colhead{Level\tablenotemark{a}} & \colhead{NIST\tablenotemark{a}} &
 \colhead{HFR\tablenotemark{b}} & \colhead{VS\tablenotemark{c}}
}
\startdata
2 & ${\rm1s2s\,^3S_{1}}$ & 7.7316132 & 7.7331 & 7.7314\\
2 & ${\rm1s2p\,^3P^o_{0}}$ & 7.7633680 & 7.7623 & 7.7631\\
2 & ${\rm1s2p\,^3P^o_{1}}$ & 7.7656952 & 7.7668 & 7.7657\\
2 & ${\rm1s2p\,^3P^o_{2}}$ & 7.7864167 & 7.7876 & 7.7862\\
2 & ${\rm1s2s\,^1S_{0}}$ & 7.7660015 & 7.7659 & 7.7658\\
2 & ${\rm1s2p\,^1P^o_{1}}$ & 7.8055847 & 7.8073 & 7.8052\\
3 & ${\rm1s2s^2\,^2S_{1/2}}$ & 7.6933430 & 7.6944 & \\
3 & ${\rm1s(^2S)2s2p(^3P^o)\,^4P^o_{1/2}}$ & 7.7071053 & 7.7063 & \\
3 & ${\rm1s(^2S)2s2p(^3P^o)\,^4P^o_{3/2}}$ & 7.7119406 & 7.7127 & \\
3 & ${\rm1s(^2S)2s2p(^3P^o)\,^4P^o_{5/2}}$ &  & 7.7291 & \\
3 & ${\rm1s(^2S)2s2p(^3P^o)\,^2P^o_{1/2}}$ & 7.7504997 & 7.7508 & \\
3 & ${\rm1s(^2S)2s2p(^3P^o)\,^2P^o_{3/2}}$ & 7.7635181 & 7.7657 & \\
3 & ${\rm1s(^2S)2p^2(^3P)\,^4P_{5/2}}$ & 7.7907946 & 7.7923 & \\
3 & ${\rm1s(^2S)2p^2(^3P)\,^4P_{3/2}}$ & 7.7836035 & 7.7841 & \\
3 & ${\rm1s(^2S)2p^2(^3P)\,^4P_{1/2}}$ & 7.7718250 & 7.7711& \\
3 & ${\rm1s(^2S)2s2p(^1P^o)\,^2P^o_{3/2}}$ & 7.7826116 & 7.7830 & \\
3 & ${\rm1s(^2S)2s2p(^1P^o)\,^2P^o_{1/2}}$ & 7.7791401 & 7.7807 & \\
3 & ${\rm1s(^2S)2p^2(^3P)\,^2P_{1/2}}$ & 7.8098881 & 7.8119& \\
3 & ${\rm1s(^2S)2p^2(^3P)\,^2P_{3/2}}$ & 7.8348090 & 7.8370& \\
3 & ${\rm1s(^2S)2p^2(^3P)\,^2D_{3/2}}$ & 7.8098881 & 7.8108& \\
3 & ${\rm1s(^2S)2p^2(^3P)\,^2D_{5/2}}$ & 7.8193109 & 7.8219& \\
3 & ${\rm1s(^2S)2p^2(^3P)\,^2S_{1/2}}$ & 7.8548944 & 7.8572& \\
\enddata
\tablenotetext{a}{NIST database \citep{nist}. Level energies are determined by
interpolation or extrapolation of known experimental values.}
\tablenotetext{b}{HFR calculations (this work).}
\tablenotetext{c}{$1/Z$ expansion calculations \citep{vai85}.}
\end{deluxetable}

\clearpage

\begin{deluxetable}{llllll}
\tabletypesize{\scriptsize}
%\rotate
\tablecaption{Wavelengths for K lines in Ni ions with
$2\leq N\leq 5$ \label{tabwl}}
\tablewidth{0pt}
\tablehead{& & & \multicolumn{3}{c}{$\lambda$ (\AA)} \\
\colhead{$N$} & \colhead{Lower level} &
 \colhead{Upper level} & \colhead{HFR\tablenotemark{a}} &
  \colhead{EXP}& \colhead{Other Theory} }
\startdata
2 & ${\rm 1s}^2\,^1{\rm S}_0$ & ${\rm 1s2p}\,^1{\rm P}^o_1$ & 1.5880 &
1.5886\tablenotemark{b} & 1.5856\tablenotemark{e} \\
 & & & & 1.5879\tablenotemark{c} & 1.5879\tablenotemark{f} \\
 & & & & 1.5886\tablenotemark{d}  & 1.58848\tablenotemark{g}\\
 & & & & & 1.5886\tablenotemark{h}\\
 2 & ${\rm 1s}^2\,^1{\rm S}_0$ & ${\rm 1s2p}\,^3{\rm P}^o_1$ & 1.5963 &
1.5966\tablenotemark{b} & 1.5941\tablenotemark{e} \\
 & & & & 1.5962\tablenotemark{c} & 1.5959\tablenotemark{f} \\
 & & & &  1.5969\tablenotemark{d} & 1.59659\tablenotemark{g}\\
 & & & & & 1.5968\tablenotemark{h}\\
 3 & ${\rm 1s}^2{\rm 2s}\,^2{\rm S}_{1/2}$ & ${\rm 1s(^2S)2s2p(^1P^o)}\,^2{\rm P}^o_{1/2}$ &
  1.5935 & 1.5940\tablenotemark{b} & 1.5911\tablenotemark{e} \\
 & & & & 1.5931\tablenotemark{c} & 1.5934\tablenotemark{f} \\
  & & & & 1.5938\tablenotemark{d} & 1.5940\tablenotemark{h} \\
3 & ${\rm 1s}^2{\rm 2s}\,^2{\rm S}_{1/2}$ & ${\rm 1s(^2S)2s2p(^3P^o)}\,^2{\rm P}^o_{3/2}$ &
  1.5966 & 1.5972\tablenotemark{b} & 1.5941\tablenotemark{e} \\
 & & & & 1.5962\tablenotemark{c} & 1.5965\tablenotemark{f} \\
 & & & & 1.5969\tablenotemark{d} &  1.5972\tablenotemark{h}\\
3 & ${\rm 1s}^2{\rm 2p}\,^2{\rm P^o}_{1/2}$ & ${\rm 1s(^2S)2p^2(^1D)}\,^2{\rm D}_{3/2}$ &
  1.5981 & 1.5987\tablenotemark{b} & 1.5956\tablenotemark{e} \\
 & & & & 1.5978\tablenotemark{c} & 1.5980\tablenotemark{f} \\
 & & & & 1.5985\tablenotemark{d} &  1.5987\tablenotemark{h}\\
3 & ${\rm 1s}^2{\rm 2s}\,^2{\rm S}_{1/2}$ & ${\rm 1s(^2S)2s2p(^3P^o)}\,^2{\rm P}^o_{1/2}$ &
  1.5996 & 1.5999\tablenotemark{b} & 1.5973\tablenotemark{e} \\
 & & & & 1.5991\tablenotemark{c} & 1.5992\tablenotemark{f} \\
 & & & & 1.5998\tablenotemark{d} &  1.6003\tablenotemark{h}\\
3 & ${\rm 1s}^2{\rm 2p}\,^2{\rm P^o}_{3/2}$ & ${\rm 1s(^2S)2p^2(^1D)}\,^2{\rm D}_{5/2}$ &
  1.6004 & 1.6011\tablenotemark{b} & 1.5983\tablenotemark{e} \\
 & & & & 1.6003\tablenotemark{c} & 1.6005\tablenotemark{f} \\
  & & & & 1.6010\tablenotemark{d} &  1.6011\tablenotemark{h}\\
3 & ${\rm 1s}^2{\rm 2p}\,^2{\rm P^o}_{3/2}$ & ${\rm 1s(^2S)2p^2(^1D)}\,^2{\rm D}_{3/2}$ &
  1.6027 & 1.6029\tablenotemark{c} & 1.6024\tablenotemark{f} \\
 & & & & 1.6036\tablenotemark{d} &  1.6033\tablenotemark{h}\\
4 & ${\rm 1s}^2{\rm 2s}^2\,^1{\rm S}_{0}$ & ${\rm 1s2s^22p}\,^1{\rm P}^o_{1}$ &
  1.6039 & 1.6046\tablenotemark{b} & 1.6015\tablenotemark{e} \\
 & & & & 1.6037\tablenotemark{c} &  1.6047\tablenotemark{h}\\
 & & & & 1.6044\tablenotemark{d} &  \\
4 & ${\rm 1s}^2{\rm 2s2p}\,^3{\rm P^o}_{2}$ & ${\rm 1s2s2p^2}\,^3{\rm D}_{3}$ &
  1.6082 & 1.6090\tablenotemark{b} & 1.6064\tablenotemark{e} \\
4 & ${\rm 1s}^2{\rm 2s2p}\,^1{\rm P^o}_{1}$ & ${\rm 1s2s2p^2}\,^1{\rm D}_{2}$ &
  1.6106 & 1.6110\tablenotemark{b} & 1.6087\tablenotemark{e} \\
5 & ${\rm 1s}^2{\rm 2s^22p}\,^2{\rm P^o}_{3/2}$ & ${\rm 1s2s^22p^2}\,^2{\rm S}_{1/2}$ &
  1.6116 & 1.6123\tablenotemark{b} & 1.6097\tablenotemark{e} \\
5 & ${\rm 1s}^2{\rm 2s^22p}\,^2{\rm P^o}_{3/2}$ & ${\rm 1s2s^22p^2}\,^2{\rm P}_{3/2}$ &
  1.6130 & 1.6135\tablenotemark{b} & 1.6109\tablenotemark{e} \\
5 & ${\rm 1s}^2{\rm 2s^22p}\,^2{\rm P^o}_{1/2}$ & ${\rm 1s2s^22p^2}\,^2{\rm D}_{3/2}$ &
  1.6137 & 1.6144\tablenotemark{b} & 1.6117\tablenotemark{e} \\
5 & ${\rm 1s}^2{\rm 2s^22p}\,^2{\rm P^o}_{1/2}$ & ${\rm 1s2s^22p^2}\,^2{\rm P}_{1/2}$ &
  1.6138 & 1.6144\tablenotemark{b} & 1.6118\tablenotemark{e} \\
5 & ${\rm 1s}^2{\rm 2s^22p}\,^2{\rm P^o}_{3/2}$ & ${\rm 1s2s^22p^2}\,^2{\rm D}_{5/2}$ &
  1.6159 & 1.6167\tablenotemark{b} & 1.6141\tablenotemark{e} \\
\enddata
%% Text for table notes should follow after the \enddata but before
%% the \end{deluxetable}. Make sure there is at least one \tablenotemark
%% in the table for each \tablenotetext.
\tablenotetext{a}{HFR calculations (this work).}
\tablenotetext{b}{JET tokamak  measurements by \cite{bom88}.The absolute accuracy is
approximately $\pm 0.001$ \AA.}
\tablenotetext{c}{TFTR tokamak  measurements by \cite{hsu87}. Normalized to
the theoretical value of \cite{vai78} for the ${\rm 1s^2\, ^1S_0 - 1s2p\,^1P_1}$ He-like transition.}
\tablenotetext{d}{TFTR tokamak  measurements by \cite{hsu87}. Normalized to
the HFS value \citep{hsu87} for the ${\rm 1s^2\, ^1S_0 - 1s2p\,^1P_1}$ He-like transition. }
\tablenotetext{e}{{\sc superstructure} calculations by \cite{bom88}.}
\tablenotetext{f}{$1/Z$ expansion calculations by \cite{vai78}.}
\tablenotetext{g}{$1/Z$ expansion calculations by \cite{vai85}.}
\tablenotetext{h}{HFS calculations by \cite{hsu87}.}
\end{deluxetable}

%\clearpage

%\begin{deluxetable}{lllll}
%\tabletypesize{\scriptsize}
%\rotate
%\tablecaption{$K\alpha_2/K\alpha_1$ line ratio for the $[1s]{\rm 3d}^M \rightarrow [2p]{\rm 3d}^M$
%transition array in Ni$^{2+}$ -- Ni$^{4+}$ \label{tabka21}}
%\tablewidth{0pt}
%\tablehead{ & \multicolumn{4}{c}{$K\alpha_2/K\alpha_1$} \\
%\colhead{$N$} & \colhead{HFR1\tablenotemark{a}} &
% \colhead{HFR2\tablenotemark{b}} & \colhead{AST1\tablenotemark{c}} &
  %\colhead{AST2\tablenotemark{d}}
%}
%\startdata
%24 & 0.50 & 0.48 & 0.50 & 0.49\\
%25 & 0.50 & 0.37 & 0.50 & 0.32\\
%26 & 0.49 & 0.49 & 0.50 & 0.50\\
%\enddata

%% Text for table notes should follow after the \enddata but before
%% the \end{deluxetable}. Make sure there is at least one \tablenotemark
%% in the table for each \tablenotetext.
%\tablenotetext{a}{{\sc hfr} calculation including the $[{\rm 3p,2p,1s}]{\rm 3d}^M$
%configurations in the CI expansion. }
%\tablenotetext{b}{{\sc hfr} calculation including the ${\{3d,4s\}}^{M-1}$ and
%the $[{\rm 3p,2p,1s}]{\rm \{3d,4s\}}^M$
%configurations in the CI expansion.}
%\tablenotetext{c}{{\sc autostructure} calculation including the $[{\rm 3p,2p,1s}]{\rm 3d}^M$
%configurations in the CI expansion. }
%\tablenotetext{d}{{\sc autostructure} calculation including the ${\{3d,4s\}}^{M-1}$ and
%the $[{\rm 3p,2p,1s}]{\rm \{3d,4s\}}^M$
%configurations in the CI expansion.}
%\end{deluxetable}

\clearpage

\begin{deluxetable}{llll}
%\tabletypesize{\scriptsize}
%\rotate
\tablecaption{$KLM$ Auger channel relative intensities in Ni$^+$ \label{tabklm}}
\tablewidth{0pt}
\tablehead{
\colhead{Ratio\tablenotemark{a}} & \colhead{EXP\tablenotemark{b}} &
 \colhead{HFR\tablenotemark{c}} & \colhead{HFR/EXP}
}
\startdata
$KL_1M_1/KLM$ & 0.076 & 0.086 & 1.13\\
$KL_1M_{23}/KLM$ & 0.152 & 0.135 & 0.89\\
$KL_{23}M_1/KLM$ & 0.136 & 0.123 & 0.90\\
$KL_{23}M_{23}/KLM$ & 0.534 & 0.653 & 1.22\\
\enddata

%% Text for table notes should follow after the \enddata but before
%% the \end{deluxetable}. Make sure there is at least one \tablenotemark
%% in the table for each \tablenotetext.
\tablenotetext{a}{The ratios are relative to the total $KLM$ Auger rate.
$K=1s$; $L_1=2s$; $L_{23}=2p$; $M_1=3s$; $M_{23}=3p$. }
\tablenotetext{b}{Solid-state measurements by \cite{egr08}. The errors are about
20\%.}
\tablenotetext{b}{HFR calculations (this work).}
\end{deluxetable}

\clearpage

\begin{deluxetable}{llll}
%\tabletypesize{\scriptsize}
%\rotate
\tablecaption{Natural $K$-level widths (eV) in singly-ionized
cosmically abundant elements \label{tabnatw}}
\tablewidth{0pt}
\tablehead{
\colhead{Element} & \colhead{REC\tablenotemark{a}} &
 \colhead{HFR}
}
\startdata
O & 0.133 & 0.129\tablenotemark{b}\\
Ne & 0.24 & 0.247\tablenotemark{c}\\
Mg & 0.334 & 0.323\tablenotemark{c}\\
Si & 0.425 & 0.407\tablenotemark{c}\\
S & 0.552 & 0.532\tablenotemark{c}\\
Ar & 0.66 & 0.643\tablenotemark{c}\\
Ca & 0.77 & 0.756\tablenotemark{c}\\
Fe & 1.19 & 1.14\tablenotemark{d}\\
Ni & 1.39 & 1.33\tablenotemark{e}\\
\enddata
\tablenotetext{a}{Recommended value by \cite{cam01}.}
\tablenotetext{b}{HFR, \cite{gar05}.}
\tablenotetext{c}{HFR, \cite{pal08}.}
\tablenotetext{d}{HFR, \cite{pal03b}.}
\tablenotetext{e}{HFR, this work.}
\end{deluxetable}

\clearpage

\begin{deluxetable}{llllllllll}
\tabletypesize{\scriptsize}
%\rotate
\tablecaption{Ni valence and K-vacancy levels with $2\leq N\leq 3$ \label{klev}}
\tablewidth{0pt}
\tablehead{
\colhead{$N$} & \colhead{$i$} &
 \colhead{$2S+1$} & \colhead{$L$} & \colhead{$2J$} &
 \colhead{Conf} & \colhead{$E$ (keV)} & \colhead{$A_r(i)$ (s$^{-1}$)} &
 \colhead{$A_a(i)$ (s$^{-1}$) } & \colhead{$\omega_K(i)$}
}
\startdata
2 & 1 & 1 & 0 & 0 & ${\rm 1s^2 \, ^1S_0}$ & 0.0000 &  &  & \\
2 & 2 & 3 & 0 & 2 & ${\rm 1s2s \, ^3S_1}$ & 7.7331 &  &  & \\
2 & 3 & 3 & 1 & 0 & ${\rm 1s2p \, ^3P^o_0}$ & 7.7623 & 3.19E+08 &  & \\
2 & 4 & 1 & 0 & 0 & ${\rm 1s2s \, ^1S_0}$ & 7.7659 &  &  & \\
2 & 5 & 3 & 1 & 2 & ${\rm 1s2p \, ^3P^o_1}$ & 7.7668 & 6.82E+13 &  & \\
2 & 6 & 3 & 1 & 4 & ${\rm 1s2p \, ^3P^o_2}$ & 7.7876 & 2.08E+09 &  & \\
2 & 7 & 1 & 1 & 2 & ${\rm 1s2p \, ^1P^o_1}$ & 7.8073 & 6.49E+14 &  & \\
3 & 1 & 2 & 0 & 1 & ${\rm 1s^22s \, ^2S_{1/2}}$ & 0.0000 &  &  & \\
3 & 2 & 2 & 1 & 1 & ${\rm 1s^22p \, ^2P^o_{1/2}}$ & 0.0528 & 2.00E+09 &  & \\
3 & 3 & 2 & 1 & 3 & ${\rm 1s^22p \, ^2P^o_{3/2}}$ & 0.0749 & 5.71E+09 &  & \\
3 & 4 & 2 & 0 & 1 & ${\rm 1s2s^2 \, ^2S_{1/2}}$ & 7.6944 & 2.63E+13 & 1.33E+14  & 0.165 \\
3 & 5 & 4 & 1 & 1 & ${\rm 1s(^2S)2s2p(^3P^o) \, ^4P^o_{1/2}}$ & 7.7063 & 7.72E+12 &
 2.25E+11  & 0.972 \\
3 & 6 & 4 & 1 & 3 & ${\rm 1s(^2S)2s2p(^3P^o) \, ^4P^o_{3/2}}$ & 7.7127 & 2.79E+13 &
 1.04E+12  & 0.964 \\
3 & 7 & 4 & 1 & 5 & ${\rm 1s(^2S)2s2p(^3P^o) \, ^4P^o_{5/2}}$ & 7.7291 & 3.91E+05 &
   & 1.000 \\
3 & 8 & 2 & 1 & 1 & ${\rm 1s(^2S)2s2p(^3P^o) \, ^2P^o_{1/2}}$ & 7.7508 & 3.73E+14 &
 3.97E+13  & 0.904 \\
3 & 9 & 2 & 1 & 3 & ${\rm 1s(^2S)2s2p(^3P^o) \, ^2P^o_{3/2}}$ & 7.7657 & 6.61E+14 &
 6.00E+09  & 1.000 \\
3 & 10 & 4 & 1 & 1 & ${\rm 1s(^2S)2p^2(^3P) \, ^4P_{1/2}}$ & 7.7711 & 4.06E+13 &
 3.56E+11  & 0.991 \\
3 & 11 & 2 & 1 & 1 & ${\rm 1s(^2S)2s2p(^3P^o) \, ^2P^o_{1/2}}$ & 7.7807 & 3.11E+14 &
 6.55E+13  & 0.826 \\
3 & 12 & 4 & 1 & 3 & ${\rm 1s(^2S)2p^2(^3P) \, ^4P_{3/2}}$ & 7.7830 & 1.47E+13 &
 1.88E+12  & 0.887 \\
3 & 13 & 2 & 1 & 3 & ${\rm 1s(^2S)2s2p(^3P^o) \, ^2P^o_{3/2}}$ & 7.7841 & 3.16E+12 &
 1.04E+14  & 0.030 \\
3 & 14 & 4 & 1 & 5 & ${\rm 1s(^2S)2p^2(^3P) \, ^4P_{5/2}}$ & 7.7923 & 6.26E+13 &
 3.05E+13  & 0.672 \\
3 & 15 & 2 & 2 & 3 & ${\rm 1s(^2S)2p^2(^1D) \, ^2D_{3/2}}$ & 7.8108 & 5.05E+14 &
 1.25E+14  & 0.802 \\
3 & 16 & 2 & 1 & 1 & ${\rm 1s(^2S)2p^2(^3P) \, ^2P_{1/2}}$ & 7.8119 & 9.53E+14 &
 1.03E+14  & 0.999 \\
3 & 17 & 2 & 2 & 5 & ${\rm 1s(^2S)2p^2(^1D) \, ^2D_{5/2}}$ & 7.8219 & 2.77E+14 &
 1.34E+14  & 0.674 \\
3 & 18 & 2 & 1 & 3 & ${\rm 1s(^2S)2p^2(^3P) \, ^2P_{3/2}}$ & 7.8370 & 8.60E+14 &
 3.72E+13  & 0.959 \\
3 & 19 & 2 & 0 & 1 & ${\rm 1s(^2S)2p^2(^1S) \, ^2S_{1/2}}$ & 7.8572 & 3.64E+14 &
 2.90E+13  & 0.926 \\
\enddata
\tablecomments{The complete version of this table is in the
electronic edition of the Journal.  The printed edition contains
only a sample.}
\end{deluxetable}

\clearpage

\begin{deluxetable}{lllllll}
\tabletypesize{\scriptsize}
%\rotate
\tablecaption{K-vacancy transitions in Ni ions with $2\leq N\leq 3$
\label{klin}}
\tablewidth{0pt}
\tablehead{
\colhead{$N$} & \colhead{$k$} &
 \colhead{$i$} & \colhead{$\lambda$} (\AA) & \colhead{$A_r(k,i)$ (s$^{-1}$)} &
 \colhead{$gf(i,k)$} & \colhead{$CF$}
}
\startdata
2 & 7 & 1 & 1.5880 & 6.49E+14 & 7.36E-01 &-1.00\\
2 & 5 & 1 & 1.5963 & 6.82E+13 & 7.82E-02 &-1.00\\
3 & 19 & 2 & 1.5886 & 8.27E+12 & 6.25E-03 &0.03\\
3 & 18 & 2 & 1.5927 & 1.15E+13 & 1.75E-02 &0.03\\
3 & 13 & 1 & 1.5928 & 3.13E+12 & 4.76E-03 &0.01\\
3 & 19 & 3 & 1.5932 & 3.55E+14 & 2.70E-01 &0.92\\
3 & 11 & 1 & 1.5935 & 3.11E+14 & 2.37E-01 &-0.46\\
3 & 9 & 1 & 1.5966 & 6.61E+14 & 1.01E+00 &-0.95\\
3 & 18 & 3 & 1.5973 & 8.49E+14 & 1.30E+00 &0.97\\
3 & 16 & 2 & 1.5979 & 7.44E+14 & 5.69E-01 &0.97\\
3 & 15 & 2 & 1.5981 & 4.46E+14 & 6.84E-01 &0.95\\
3 & 8 & 1 & 1.5996 & 3.73E+14 & 2.86E-01 &-0.94\\
3 & 17 & 3 & 1.6004 & 2.77E+14 & 6.38E-01 &0.95\\
3 & 16 & 3 & 1.6025 & 2.09E+14 & 1.61E-01 &-0.47\\
3 & 15 & 3 & 1.6027 & 5.88E+13 & 9.06E-02 &0.13\\
3 & 12 & 2 & 1.6039 & 1.94E+11 & 2.99E-04 &-0.02\\
3 & 10 & 2 & 1.6064 & 4.04E+13 & 3.13E-02 &0.94\\
3 & 14 & 3 & 1.6066 & 6.26E+13 & 1.45E-01 &-0.95\\
3 & 6 & 1 & 1.6075 & 2.79E+13 & 4.33E-02 &-0.96\\
3 & 12 & 3 & 1.6085 & 1.46E+13 & 2.26E-02 &0.97\\
3 & 5 & 1 & 1.6089 & 7.72E+12 & 6.00E-03 &-0.96\\
3 & 10 & 3 & 1.6110 & 2.08E+11 & 1.61E-04 &0.01\\
3 & 4 & 2 & 1.6225 & 1.38E+13 & 1.09E-02 &0.76\\
3 & 4 & 3 & 1.6272 & 1.24E+13 & 9.86E-03 &0.42\\
\enddata
\tablecomments{The complete version of this table is in the
electronic edition of the Journal.  The printed edition contains
only a sample.}
\end{deluxetable}

%% Tables may also be prepared as separate files. See the accompanying
%% sample file table.tex for an example of an external table file.
%% To include an external file in your main document, use the \input
%% command. Uncomment the line below to include table.tex in this
%% sample file. (Note that you will need to comment out the \documentclass,
%% \begin{document}, and \end{document} commands from table.tex if you want
%% to include it in this document.)


%% The following command ends your manuscript. LaTeX will ignore any text

\end{document}